\documentclass[sigconf]{acmart}

\usepackage{booktabs} 

\usepackage{amssymb}
\usepackage{amsthm, amsmath}
\usepackage{algorithmic}
\usepackage[ruled]{algorithm2e}
\usepackage{color}
\usepackage{url}
\usepackage{xfrac}

\usepackage{subfigure}
\usepackage{tablefootnote}

\begin{document}
\title{Growing Story Forest Online from Massive Breaking News}

\copyrightyear{2017} 
\acmYear{2017} 
\setcopyright{acmcopyright}
\acmConference{CIKM'17}{}{November 6--10, 2017, Singapore.}
\acmPrice{15.00}
\acmDOI{https://doi.org/10.1145/3132847.3132852}
\acmISBN{ISBN 978-1-4503-4918-5/17/11}

\fancyhead{}
\settopmatter{printacmref=false, printfolios=false}

\author{Bang Liu$^1$, Di Niu$^1$, Kunfeng Lai$^2$, Linglong Kong$^1$, Yu Xu$^2$}
       \affiliation{$^1$University of Alberta, Edmonton, AB, Canada}
 \affiliation{$^2$Mobile Internet Group, Tencent Inc., Shenzhen, China}



\begin{abstract}
We describe our experience of implementing a news content organization system at Tencent that discovers events from vast streams of breaking news and evolves news story structures in an online fashion.  
Our real-world system has distinct requirements in contrast to previous studies on topic detection and tracking (TDT) and event timeline or graph generation, in that we 1) need to accurately and quickly extract distinguishable events from massive streams of long text documents that cover diverse topics and contain highly redundant information, and 2) must develop the structures of event stories in an online manner, without repeatedly restructuring previously formed stories, in order to guarantee a consistent user viewing experience. In solving these challenges, we propose \emph{Story Forest}, a set of online schemes that automatically clusters streaming documents into events, while connecting related events in growing trees to tell evolving stories. 
We conducted extensive evaluation based on 60 GB of real-world Chinese news data, although our ideas are not language-dependent and can easily be extended to other languages, through detailed pilot user experience studies. The results demonstrate the superior capability of Story Forest to accurately identify events and organize news text into a logical structure that is appealing to human readers, compared to multiple existing algorithm frameworks.\footnote{This work was sponsored by CCF-Tencent Open Fund.}

\end{abstract}

%
%



\maketitle

\section{Introduction}
\label{sec:intro}

With information explosion in the fast-paced modern society, tremendous volumes of articles on trending and breaking news are being generated on a daily basis by various Internet media providers, e.g., Yahoo! News, CNN, Tencent News, Sina News, etc. In the meantime, it becomes increasingly difficult for normal readers to digest such a large amount of streaming news information. Search engines perform document retrieval from large corpora based on user-defined queries that specify what are interesting to the user. However, they do not provide a natural way for users to view what is going on. Furthermore, search engines return a list of ranked documents and do not provide structural summaries of trending topics or breaking news.

An emerging alternative way to visualize news corpora without pre-specified queries is to organize and present news articles through event timelines \cite{yan2011evolutionary,  wang2016socially}, event threads \cite{nallapati2004event}, event evolution graphs \cite{yang2009discovering}, or information maps \cite{shahaf2012trains,shahaf2013information,xu2013summarizing}. However, till today few existing news information organization techniques are turned into large-scale deployment due to several reasons:

First of all, despite research efforts in Topic Detection and Tracking (TDT) \cite{yang2002multi, allan1998line}, it remains challenging to extract distinguishable ``events'' at a proper granularity, as building blocks of the news graph, from today's vast amount of open-domain daily news. The news articles may cover extremely diverse topics and contain redundant information about a same conceptual event published by different sources. For example, simply connecting individual articles \cite{shahaf2012trains} or named entities \cite{faloutsos2004fast} in a graph will lead to redundant and entangled information. On the other hand, connecting co-occuring keyword sets in an information map \cite{shahaf2013information} can greatly reduce the fine details of news graphs. But even with the keyword graph, a user still needs to put additional efforts to understand the large number of articles associated with each keyword set.    

Second, many recently proposed event graphs or information maps try to link events in an arbitrary evolution graph \cite{yang2009discovering} or permitting intertwining branches in the information map \cite{shahaf2013information}. However, we would like to point out that such overly complex graph structures do not make it easy for users to quickly visualize and understand news data. In fact, unlike a novel or a complex story about a celebrity queried from a search engine, most breaking news stories follow one of a few typical developing structures. In fact, for breaking news summary that will appeal to commercial uses, simple story structures are preferred.

Most importantly, most existing event timeline or event graph generation schemes are based on \emph{offline} optimization over the entire news corpora, while for a system that visualizes breaking news, it is desirable to ``grow'' the stories in an online fashion without disrupting or restructuring the previously generated stories. On one hand, online computation can prevent repeated processing of older documents. On the other hand, an online scheme can deliver a consistent story development structure to users, so that users can quickly visualize what's new in the hot events that they are trying to follow. Furthermore, given the vast amount of collected daily news data, the entire online computation to identify new events and extend the existing story graphs will incur less delay.


In this paper, we present our experience of implementing \textit{StoryForest}, which is a comprehensive system to organize vast amounts of breaking news data into easily readable story trees of events in an online fashion.
We make careful design choices for each component in this large system, with the following contributions:

First, our system can accurately cluster massive amounts of long news documents into conceptually clean events through a novel two-layer document clustering procedure that leverages a wide range of feature engineering and machine learning techniques, mainly including keyword extraction, keyword community detection, a pre-trained classifier to detect whether two documents are talking about the same event, and a graph-based document clustering procedure. On a labeled news dataset, our proposed text clustering procedure significantly outperforms a number of existing text clustering schemes.

Second, our system further groups the discovered events into stories, where each story is represented by a \emph{tree} of events. A link between two events indicates the temporal migration or a causal relationship between two events. Compared with existing story generation systems such as StoryGraph \cite{yang2009discovering} and MetroMap \cite{shahaf2012trains}, we propose an online algorithm to evolve story trees incrementally based on daily news, without any churn of reforming the graph when new data arrive.
As a result, each story is presented in one of several easy-to-view structures, i.e., either a linear timeline, a flat structure, or a tree with branches, which we believe are sufficient to represent story structures of most breaking news. 

Finally, we evaluated the performance of our system based on 60 GB of Chinese news documents collected from all the major Internet news providers in China (including Tencent, Sina, WeChat, Sohu, etc.) in a three-month period from October 1, 2016 to December 31, 2016, covering extremely diverse topics in the open domain. 
We also conducted a detailed and extensive pilot user experience study for (long) news document clustering and news story generation to evaluate how our system as well as several baseline schemes conform to the habit of human readers.

According to the pilot user experience study, our system outperforms multiple state-of-the-art news clustering and story structure generation systems such as KeyGraph \cite{sayyadi2013graph} and StoryGraph \cite{yang2009discovering} in terms of logical validity of the generated story structures, as well as the conceptual purity of each identified event and story. Experiments show that the average time for our Java-based system to finish event clustering and story structure generation based on the daily data is less than $30$ seconds on  a MacBook Pro with a 2 GHz Intel Core i7 processor, and 8 GB memory. Therefore, our system proves to be highly efficient and practical.


It is worth mentioning that our work represents the first system that is able to efficiently process vast amounts of Chinese news data into organized story structures, although our proposed algorithms and schemes are also applicable to news data in English (and other languages) by simply replacing the word segmentation and NLP tools with the counterparts for the corresponding language.    

\begin{figure*}
\includegraphics[width=6.7in]{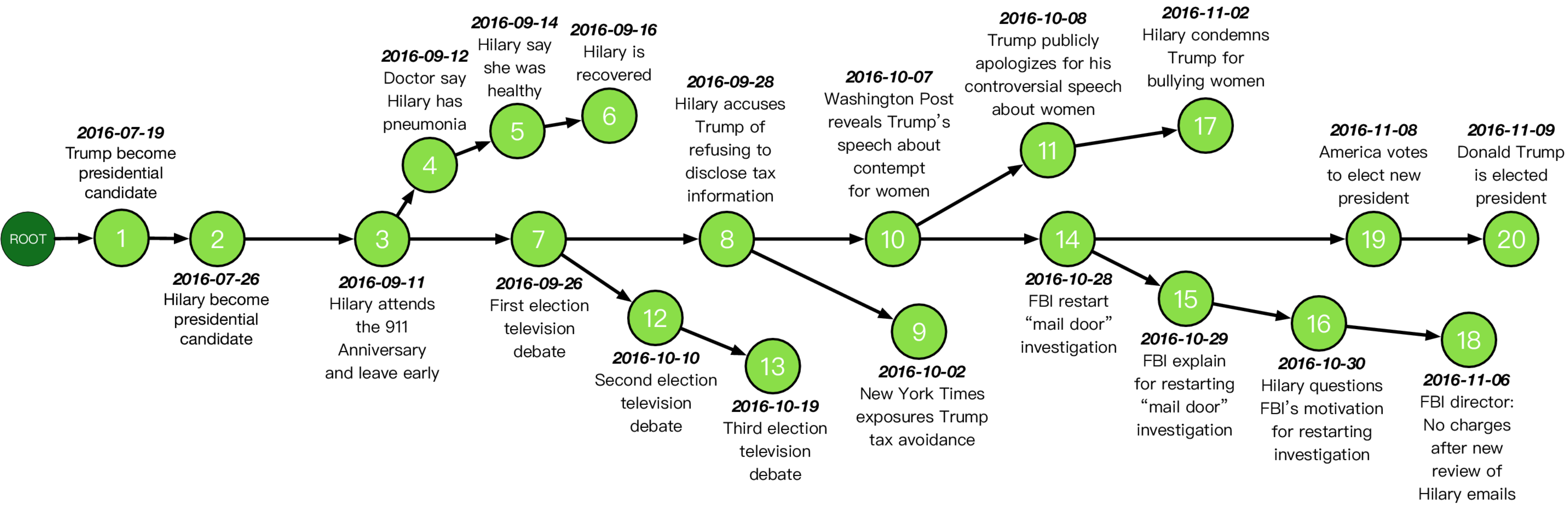}
\caption{The story tree of ``2016 U.S. presidential election.''}
\label{fig:CaseStudy}
\vspace{-2mm}
\end{figure*}

\section{Problem Definition and Notations}
\label{sec:problem}



We first present some definitions of key concepts in the top-down hierarchy: \textit{topic} $\rightarrow$ \textit{story} $\rightarrow$ \textit{event} to be used in this paper.

\begin{definition}
  \textit{Event}: an event $\mathcal{E}$ is a set of one or several documents that contain highly similar information.
\end{definition}

\begin{definition}
  \textit{Story}: a story $\mathcal{S}$ is a tree of events that revolve around a group of specific persons and happen at certain places during specific times. A directed edge from event $\mathcal{E}_1$ to $\mathcal{E}_2$ indicates a temporal evolution or a logical connection from $\mathcal{E}_1$ to $\mathcal{E}_2$.
\end{definition}

\begin{definition}
  \textit{Topic}: a topic consists of a set of stories that are highly correlated or similar to each other.
  \vspace{-1mm}
\end{definition}

Each topic may contain multiple story trees, and each story tree consists of multiple logically connected events.
In our work, events (instead of news documents) are the smallest atomic units. Each event is also assumed to belong to a single story and contains partial information about that story.
For instance, considering the topic \textit{American presidential election}, \textit{2016 U.S. presidential election} is a story within this topic, and  \textit{Trump and Hilary's first television debate} is an event within this story.

We now introduce some notations and describe our problem formally. Given a news document stream $D = \{ \mathcal{D}_1, \mathcal{D}_2, \ldots, \mathcal{D}_t,\ldots \}$, where $\mathcal{D}_t$ is the set of news documents collected on time period $t$, our objective is to: a) cluster all news documents $D$ into a set of events $E = \{ \mathcal{E}_1, \ldots, \mathcal{E}_{|E|} \}$, and b) connect the extracted events to form a set of stories $S = \{ \mathcal{S}_1, ..., \mathcal{S}_{|S|} \}$. Each story $\mathcal{S} = (E, L)$ contains a set of events $E$ and a set of links $L$, where $L_{i,j} := <\mathcal{E}_i, \mathcal{E}_j>$ denotes a directed link from event $\mathcal{E}_i$ to $\mathcal{E}_j$, which indicates a temporal evolution or logical connection relationship.

Furthermore, we require the events and story trees to be extracted in an online or incremental manner. That is, we extract events from each $\mathcal D_t$ individually when the news corpus $\mathcal D_t$ arrives in time period $t$, and \emph{merge} the discovered events into the existing story trees that were found at time $t-1$. This is a unique strength of our scheme as compared to prior work, since we do not need to repeatedly process older documents and can deliver  a set of evolving yet logically consistent story trees to users.  


\begin{figure}
\includegraphics[width=3.4in]{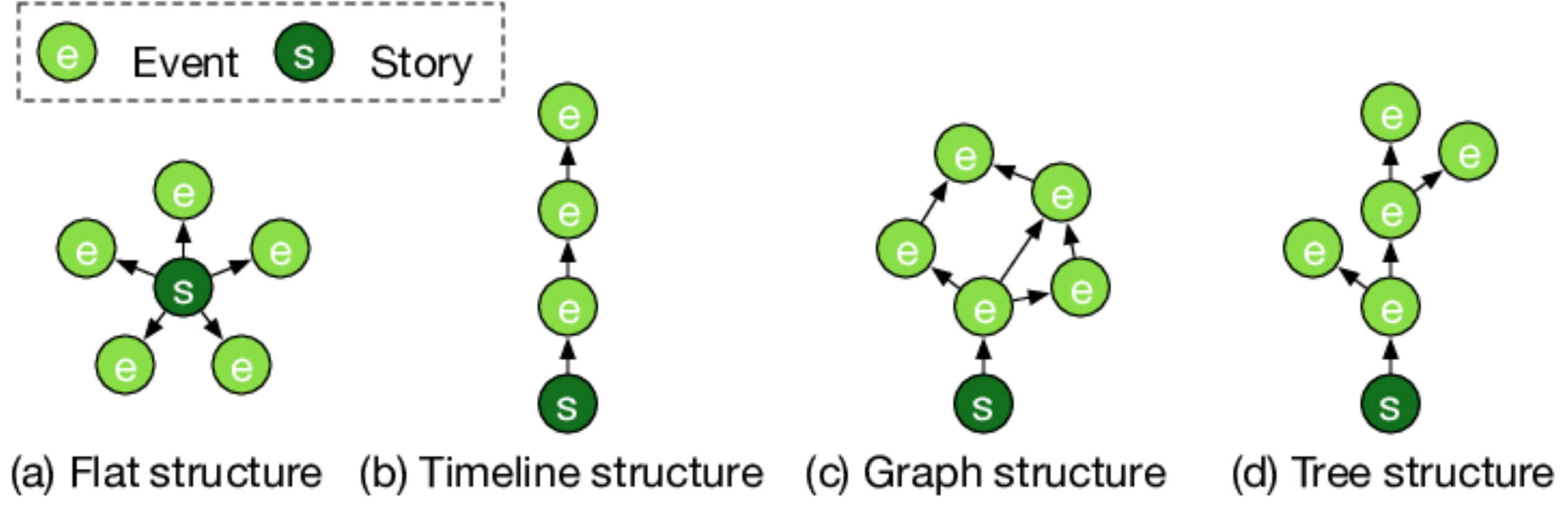}
\caption{Different structures to characterize a story.}
\vspace{-2mm}
\label{fig:storyStructures}
\vspace{-2mm}
\end{figure}

For example, Fig.~\ref{fig:CaseStudy} illustrates the story tree of ``2016 U.S. presidential election''. The story contains $20$ nodes, where each node indicates an event in 2016 U.S. election, and each link indicates a temporal evolution or a logical connection between two events. 
The index number on each node represents the event sequence over the timeline. There are $6$ paths within this story tree, where the path $1 \rightarrow 20$ indicates the whole presidential election process, branch $3 \rightarrow 6$ is about Hilary's health conditions, branch $7 \rightarrow 13$ talks about television debates, $14 \rightarrow 18$ depicts the investigation into Hilary's ``mail door'', etc. As we can see, by modeling the evolutionary and logical structure of a story into a story tree, users can easily grasp the logic of news stories and learn the main information quickly.

Let us represent each story by an empty root node $s$ from which the story is originated, and denote each event by an event node $e$. The events in a story can be organized in one of the following four structures shown in Fig. \ref{fig:storyStructures}: a) a flat structure that does not include dependencies between events; b) a timeline structure that organizes events by their timestamps; c) a graph structure that checks the connection between all pairs of events and maintains a subset of most strong connections; d) a tree structure, which represents a story's evolving structure by a tree.  

Compared with a tree structure, sorting events by timestamps omits the logical connection between events, while using directed acyclic graphs to model event dependencies without considering the evolving consistency of the whole story can leads to unnecessary connections between events.
Through extensive user experience studies in Sec.~\ref{sec:eval}, we show that tree structures are the most effective way to represent breaking news stories as compared to other structures, including the more complex graph structures. 

\section{The Story Forest System}
\label{sec:system}

\begin{figure*}
\includegraphics[width=6.7in]{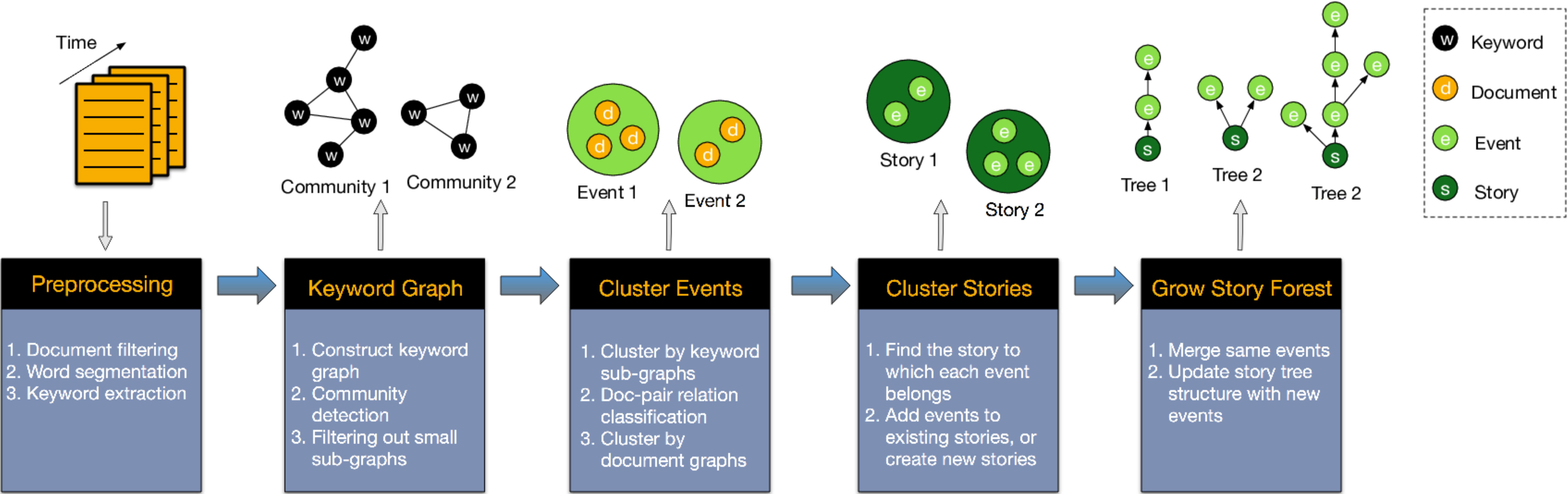}
\caption{An overview of the system architecture of \textit{Story Forest}.}
\label{fig:system}
\vspace{-0mm}
\end{figure*}



An overview of our \textit{Story Forest} system is shown in Fig.~\ref{fig:system}, which mainly consists of three components: preprocessing, document clustering and story tree update, divided into 5 steps. First, the input news document stream will be processed by a variety of NLP and machine learning tools, mainly including document filtering, word segmentation and keyword extraction. Second, steps 2--3 will cluster documents into events in a novel 2-layer procedure as follows.
For news corpus $\mathcal D_t$ in each time period $t$, we form a keyword graph \cite{sayyadi2013graph} from these documents based on keyword co-occurrence, and extract topics as subgraphs from the keyword graph using community detection algorithms. The topics with few keywords will be discarded. After each topic is found, we find all the documents associated with the topic, and further cluster these documents into events through a semi-supervised document clustering procedure aided by a pre-trained document-pair relationship classifier.
Finally, in steps 4--5 we update the story trees (formed previously) by either inserting each discovered event into an existing story tree at the right place, or creating a new story tree if the event does not belong to any existing story. Note that each topic may contain multiple story trees and each story tree consists of logically connected events.
We will explain the design choices of each component in detail in the following.

\subsection{Preprocessing}
\label{subsec:preprocessing}
When a new set of news documents arrives,  we need to clean, filter documents, and extract features that will be helpful to the steps that follow. 
Our preprocessing module mainly includes the following three steps, which are critical to the overall system performance:

\textbf{Document filtering}: unimportant documents with content length smaller than a threshold (20 characters) will be discarded.

\textbf{Word segmentation}: we segment the title and body of each document using Stanford Chinese Word Segmenter \textit{Version 3.6.0} \cite{chang2008optimizing}, which has proved to yield excellent performance on Chinese word segmentation tasks. Note that for data in a different language, the corresponding word segmentation tool in that language can be used instead. 


\textbf{Keyword extraction}: extracting keywords from each document to represent the main concepts of the document is quite critical to the performance and efficiency of the entire system. We found that traditional keyword extraction approaches, such as TF-IDF based keyword extraction and TextRank \cite{mihalcea2004textrank}, are not sufficient to achieve good performance for real-world news data. For example, the TF-IDF based method measures each word's importance by frequency information; it cannot detect keywords that yet have a relatively low frequency. The TextRank algorithm utilizes the word co-occurrence information and is able to handle such cases. However, its efficiency is relatively low, with time cost increasing significantly as the document length increases.

\begin{table}
  \caption{Features for the classifier to extract keywords.}
  \label{tab:features}
  \begin{tabular}{lp{5.5cm}}
    \toprule
    Type & Features\\
    \midrule
    Word feature & Named entity or not, location name or not, contains angle brackets or not. \\
    Structural feature & TF-IDF, whether appear in title, first occurrence position in document, average occurrence position in document, distance between first and last occurrence positions, average distance between word adjacent occurrences, percentage of sentences that contains the word, TextRank score.\\
    Semantic feature & LDA\tablefootnote{We trained a $1000$-dimensional LDA model based on news data collected from January 1, 2016 to May 31, 2016 that contains $300,000+$ documents.
    }\\
    \bottomrule
  \end{tabular}
  \vspace{-3mm}
\end{table}

\begin{figure}
\includegraphics[width=3.0in]{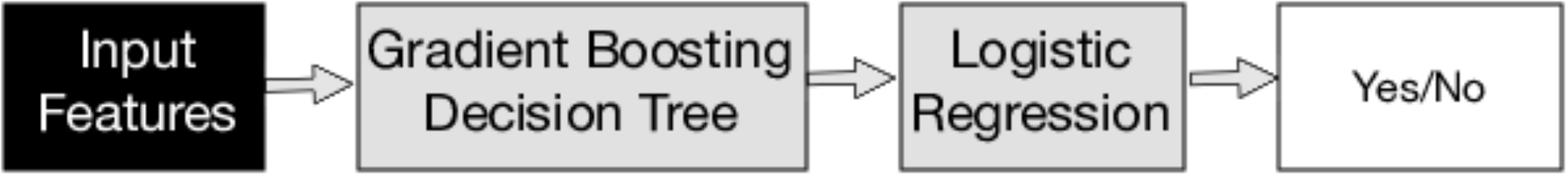}
\caption{The classifier to extract keywords.}
\vspace{-1mm}
\label{fig:keywordClassify}
\vspace{-3mm}
\end{figure}

To efficiently and accurately extract keywords, we constructed a supervised learning system to classify whether a word is a keyword or not for a document.
In particular, we manually labeled the keywords of $10,000+$ documents, including $20,000+$ positive keyword samples and $350,000+$ negative samples.
Table~\ref{tab:features} lists the main features that we found critical to the binary classifier.

A straightforward idea is to input the raw features listed above to a Logistic Regression (LR). However, as a linear classifier, LR relies on careful feature engineering.
To reduce the impact of human judgement in feature engineering, we combine a Gradient Boosting Decision Tree (GBDT) with the LR classifier to get the binary yes/no classification result, as shown in Fig. \ref{fig:keywordClassify}. GBDT, as a nonlinear model, can automatically discover useful cross features or feature combinations from raw features and discretize continuous features. 
The output of the GBDT will serve as the input of the LR classifier. Finally, the LR classifier will determine whether a word is a keyword or not for the document in question. We also tried SVM as the classifier in the second layer instead of LR and observed similar performance. Our final keyword extraction precision and recall rate are $0.83$ and $0.76$, while they are $0.72$ and $0.76$ respectively if we don't add the GBDT component.

\subsection{Document Clustering and Event Extraction}
\label{subsec:eventClustering}

After document preprocessing, we need to extract events. Event extraction here is essentially a fine-tuned document clustering procedure to group conceptually similar documents into events. Although clustering studies are often subjective in nature, we show that our carefully designed procedure can significantly improve the accuracy of event clustering, conforming to human understanding, based on a manually labeled news dataset.
To handle the high accuracy requirement for long news text clustering, we propose a $2$-layer clustering approach based on both keyword graphs and document graphs.

\textit{First}, we construct a large keyword co-occurrence graph \cite{sayyadi2013graph} $\mathcal{G}$. Each node in $\mathcal{G}$ is a keyword $w$ extracted by the scheme described in Sec.~\ref{subsec:preprocessing}, and each undirected edge $e_{i,j}$ indicates that $w_i$ and $w_j$ have ever co-occured in a same document. 
Edges that satisfy two conditions will be kept and other edges will be dropped: the times of co-occurrence shall be above a minimum threshold (we use $3$ in our system), and the conditional probabilities of the occurrence $\Pr\{w_j| w_i\}$ and $\Pr\{w_i | w_j\}$ also need to be bigger than a predefined threshold (we use $0.15$), where the conditional probability $\Pr\{w_j| w_i\}$ represents the probability that $w_j$ occurs in a document if the document contains word $w_i$.

\textit{Second}, we perform community detection in the constructed keyword graph. This step aims to split the whole keyword graph $\mathcal{G}$ into communities $C = \{\mathcal{C}_1, \mathcal{C}_1, ..., \mathcal{C}_{|C|}\}$, where each community $\mathcal{C}_i$ contains the keywords for a certain topic (to which multiple stories may be associated). 
The benefit of using community detection in the keyword graph is that each keyword can appear in multiple communities, which makes sense in reality. 
We also tried another method of clustering keywords by \textit{Word2Vec}.
However, the performance is worse than community detection based on co-occurrence graphs. The reason is that using word vectors tends to cluster the words with similar semantic meanings. However, unlike articles in a specialized domain, in long news documents in the open domain, it is highly possible that keywords with different semantic meanings can co-occur in the same event.

To detect keyword communities, we utilize the \emph{betweenness centrality score} \cite{sayyadi2013graph} of edges to measure the strength of each edge in the keyword graph. An edge's betweenness score is defined as the number of shortest paths between all pairs of nodes that pass through it. An edge between two communities is expected to achieve a high betweenness score. Edges with high betweenness score will be removed iteratively to extract communities. The iterative splitting process will stop until the number of nodes in each sub-graph is smaller than a predefined threshold, or until the maximum betweenness score of all edges in the sub-graph is smaller than a threshold that depends on the sub-graph's size. We refer interested readers to \cite{sayyadi2013graph} for more details about community detection.

After we obtain the keyword communities, we calculate the cosine similarity between each document and a keyword community.  The documents are represented by TF-IDF vectors. As a keyword community is a bag of words, it can also be considered as a document. We assign each document to the keyword community which gives the highest similarity and the similarity is above a predefined threshold. Up to now, we have finished document clustering in the first layer, i.e., the documents are grouped according to topics. 

\textit{Third}, we further perform the second-layer document clustering within each topic to obtain fine-grained events. We also call this process \emph{event clustering}. An event only contains documents that talk about the same semantic event. To yield fine-grained event clustering, unsupervised learning is not sufficient. 
Instead, we adopt a supervised-learning-guided clustering procedure in the second layer.

Specifically, we train an SVM classifier to determine whether a pair of documents are talking about the same event or not using a bunch of document-pair features as the input, including the cosine similarities of content TF-IDF and TF vectors, the cosine similarities of title TF-IDF and TF vectors, the similarity of the first sentences in the two documents, etc.

For each pair of documents within a same topic, we decide whether to connect them or not according to the prediction made by the document-pair relationship classifier mentioned above. Hence, the documents in each topic will form a document graph. We then apply the same community detection algorithm mention above to such document graphs. 
Note that the graph-based clustering on the second layer is highly efficient, since the number of documents contained in each topic is significantly smaller after the first-layer document clustering. 

In a nutshell, our 2-layer scheme groups documents into topics based on keyword community detection and further groups the documents within each topic into fine-grained events. For each event $\mathcal{E}$, we also record the set of keywords $\mathcal{C}_{\mathcal{E}}$ of the topic (keyword community) which it belongs to, which will be helpful in the subsequent story tree development.

\subsection{Growing Story Trees Online}
\label{sec:tree}

Given the set of extracted events for a particular topic, we further organize these events into multiple stories under this topic in an online manner. Each story is represented by a \textit{Story Tree} to characterize the evolving structure of that story.
Upon the arrival of a new event and given an existing story forest, our online algorithm to grow the story forest mainly involves two steps: a) identifying the story tree to which the event belongs; b) updating the found story tree by inserting the new event at the right place. 
If this event does not belong to any existing story, we create a new story tree.

{\bf a) Identifying the related story tree.} 
Given a set of new events $E_t = \{\mathcal{E}_1, \mathcal{E}_2, ..., \mathcal{E}_{|E_t|}\}$ at time period $t$ and an existing story forest $\mathcal{F}_{t-1} = \{ \mathcal{S}_1, \mathcal{S}_2, ..., \mathcal{S}_{|\mathcal{F}_{t-1}|}\}$ that has been formed during previous $t-1$ time periods, our objective is to assign each new event $\mathcal{E} \in E_t$ to an existing story tree $\mathcal{S} \in \mathcal{F}_{t-1}$. If no story in the current story forest matches that event, a new story tree will be created and added to the story forest. 

We apply a two-step strategy to decide whether a new event $\mathcal{E}$ belongs to an existing story tree $\mathcal{S}$ formed previously.
First, as described at the end of Sec. \ref{subsec:eventClustering}, event $\mathcal{E}$ has its own keyword set $\mathcal{C}_{\mathcal{E}}$.
Similarly, for the existing story tree $\mathcal{S}$, there is an associated keyword set $\mathcal{C}_{\mathcal{S}}$ that is a union of all the keyword sets of the events in that tree.

Then, we can calculate the compatibility between event $\mathcal{E}$ and story tree $\mathcal{S}$ as the Jaccard similarity coefficient between $\mathcal{C}_{\mathcal{S}}$ and $\mathcal{C}_{\mathcal{E}}$: 
$
  \text{compatibility}(\mathcal{C}_{\mathcal{S}}, \mathcal{C}_{\mathcal{E}}) = \frac{|\mathcal{C}_{\mathcal{S}} \cap \mathcal{C}_{\mathcal{E}}|}{|\mathcal{C}_{\mathcal{S}} \cup \mathcal{C}_{\mathcal{E}}|}.
$
If the compatibility is bigger than a threshold, we further check whether at least a document in event $\mathcal{E}$ and at least a document in story tree $\mathcal{S}$ share $n$ or more common words in their titles (with stop words removed). If yes, we assign event $\mathcal{E}$ to story tree $\mathcal{S}$. Otherwise, they are not related. In our experiments, we set $n=1$. 
If the event $\mathcal{E}$ is not related to any existing story tree, a new story tree will be created.

\begin{figure}
\includegraphics[width=3.3in]{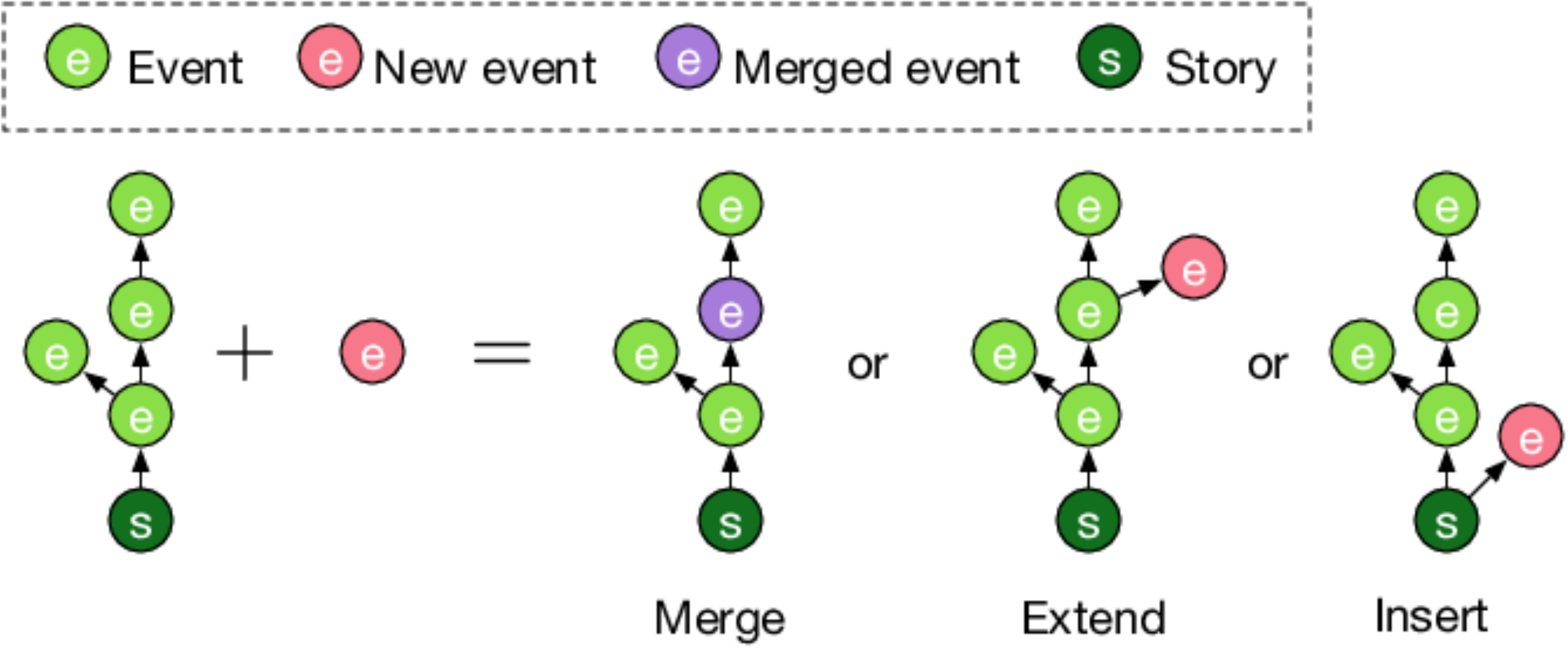}
\caption{Three types of operations to place a new event into its related story tree.}
\label{fig:nodeOperations}
\vspace{-2mm}
\end{figure}

{\bf b) Updating the related story tree.} After a related story tree $\mathcal{S}$ has been identified for the incoming event $\mathcal{E}$, we perform one of the 3 types of operations to place event $\mathcal{E}$ in the tree: \textit{merge}, \textit{extend} or \textit{insert}, as shown in Fig.~\ref{fig:nodeOperations}.
The \textit{merge} operation merges the new event $\mathcal{E}$ into an existing event node in the tree. The \textit{extend} operation will append event $\mathcal{E}$ as a child node to an existing event node in the tree. Finally, the \textit{insert} operation directly appends event $\mathcal{E}$ to the root node of story tree $\mathcal{S}$. Our system chooses the most appropriate operation to process the incoming event based on the following procedures.

{\bf \emph{Merge}}: we merge $\mathcal{E}$ with an existing event in the tree, if they essentially talk about the same event.
This can be achieved by checking whether the centroid documents of the two events are talking about the same thing using the document-pair relationship classifier described in Sec.~\ref{subsec:eventClustering}. The centroid document of an event is simply the concatenation of all the documents in the event.

{\bf\emph{Extend}} \emph{and} {\bf \emph{Insert}}: if event $\mathcal{E}$ does not overlap with any existing event, we will find the parent event node in $\mathcal{S}$ to which it should be appended.
We calculate the \emph{connection strength} between the new event $\mathcal{E}$ and each existing event $\mathcal{E}_j \in \mathcal{S}$ based on three factors: 1) the time distance between $\mathcal{E}$ and $\mathcal{E}_j$, 2) the compatibility of the two events, and 3) the \emph{storyline coherence} if $\mathcal{E}$ is appended to $\mathcal{E}_j$ in the tree, i.e., 
\begin{align}
\label{eqn:linkScore} 
\begin{split}
  \text{ConnectionStrength}(\mathcal{E}_j, \mathcal{E})  :=\ \text{compatibility}(\mathcal{E}_j, \mathcal{E}) \times \\
  \text{coherence}(\mathcal{L}_{\mathcal{S}\to\mathcal{E}_j\to \mathcal{E}}) \times \text{timePenalty}(\mathcal{E}_j, \mathcal{E}).
\end{split}
\end{align}

Now we explain the three components in the above equation one by one. \emph{First}, the compatibility between two events $\mathcal{E}_i$ and $\mathcal{E}_j$ is given by
\begin{equation}
  \text{compatibility}(\mathcal{E}_i, \mathcal{E}_j) = \frac{\text{TF}(d_{c_i}) \cdot \text{TF}(d_{c_{j}})}{\|\text{TF}(d_{c_i})\| \cdot \|\text{TF}(d_{c_{j}})\|},
\end{equation}
where $d_{c_i}$ is the centroid document of event $\mathcal{E}_i$.

Furthermore, the storyline of $\mathcal{E}_j$ is defined as the path in $\mathcal{S}$ starting from the root node of $\mathcal{S}$ ending at $\mathcal{E}_j$ itself, denoted by $\mathcal{L}_{\mathcal{S}\rightarrow \mathcal{E}_j}$. Similarly, the storyline of $\mathcal{E}$ appended to $\mathcal{E}_j$ is denoted by $\mathcal{L}_{\mathcal{S}\rightarrow \mathcal{E}_j\rightarrow\mathcal{E}}$.
For a storyline $\mathcal{L}$ represented by a path
$\mathcal{E}^0\to \ldots \to \mathcal{E}^{|\mathcal{L}|}$, where $\mathcal{E}^0 := \mathcal S$, its \textit{coherence} \cite{xu2013summarizing} measures the theme consistency along the storyline, and is defined as
\begin{equation}
  \text{coherence}(\mathcal{L}) = \frac{1}{|\mathcal{L}|}\sum_{i=0}^{|\mathcal{L}|-1} \text{compatibility}(\mathcal{E}^i, \mathcal{E}^{i+1}),
\end{equation}

Finally, the bigger the time gap between two events, the less possible that the two events are connected. We thus calculate time penalty by
\begin{align}
  \text{timePenalty}(\mathcal{E}_j, \mathcal{E}) = \begin{cases}
  e^{\delta \cdot (t_{\mathcal{E}_j} - t_{\mathcal{E}})}
   &\ \text{if } t_{\mathcal{E}_j} - t_{\mathcal{E}} < 0\\
  0 &\ \text{otherwise}\\
  \end{cases}
\end{align}
where $t_{\mathcal{E}_j}$ and  $t_{\mathcal{E}}$ are the timestamps of event $\mathcal{E}_j$ and $\mathcal{E}$ respectively. The timestamp of an event is the minimum timestamp of all the documents in the event.

We calculate the connection strength between the new event $\mathcal{E}$ and every event node $\mathcal{E}_j \in \mathcal{S}$ using \eqref{eqn:linkScore}, and append event $\mathcal{E}$ to the existing $\mathcal{E}_j$ that leads to the maximum connection strength. 
If the maximum connection strength is lower than a threshold value, we \textit{insert} $\mathcal{E}$ into story tree $\mathcal{S}$ by directly appending it to the root node of $\mathcal{S}$. In other words, \emph{insert} is a special case of \emph{extend}.

\section{Evaluation}
\label{sec:eval}

\begin{figure}[t]
		\hspace{7mm}
        \includegraphics[width=3.2in]{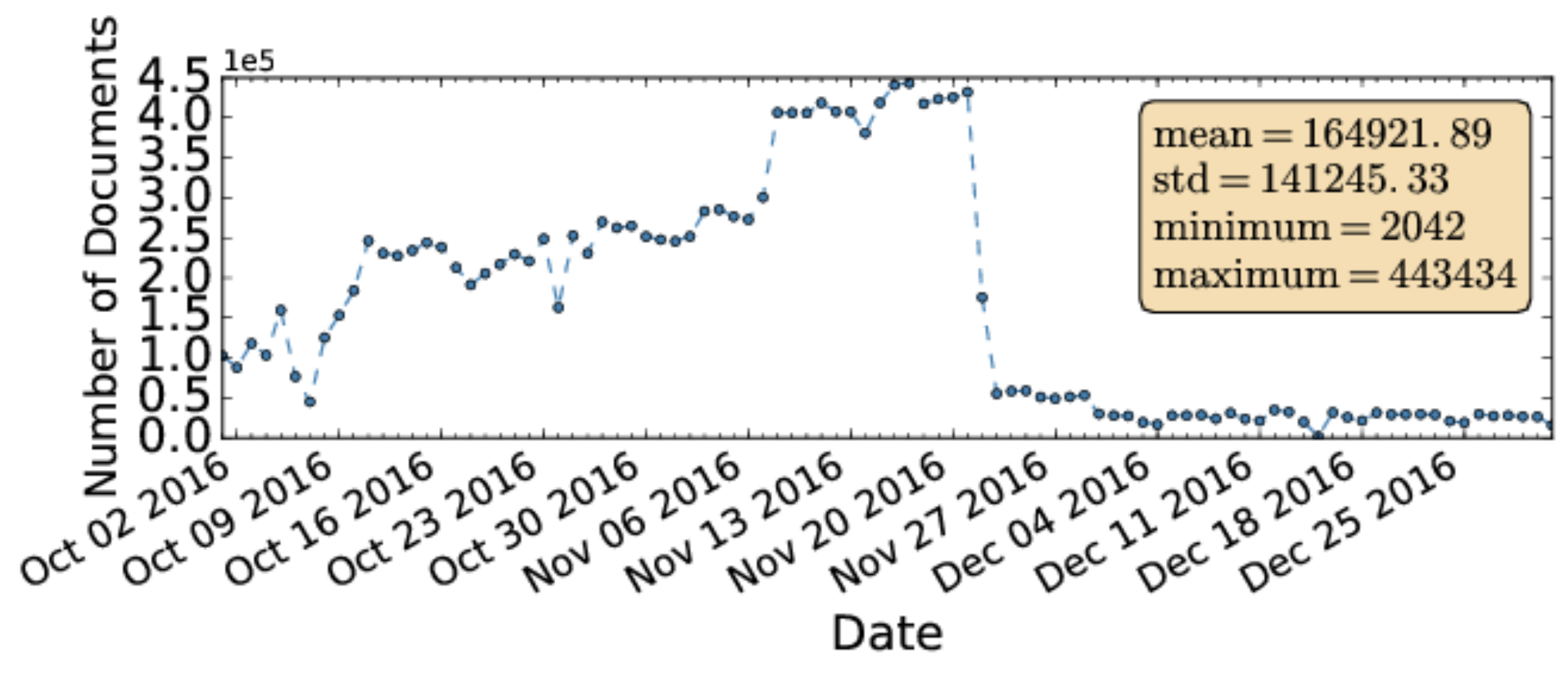}
        \vspace{-5mm}
        \caption{The number of documents on different days in the dataset.}
        \label{fig:docAmount}
        \vspace{-5mm}
\end{figure}


We evaluate the performance of our system based on 60 GB of Chinese news documents collected from all the major Internet news providers in China, such as Tencent and Sina, in a three-month period from October 1, 2016 to December 31, 2016 covering different topics in the open domain. Fig.~\ref{fig:docAmount} shows the amounts of documents on different days in the dataset. The average number of documents in one day during that period is $164,922$. For the following experiments, we use the data in the first $7$ days for parameter tuning. The remaining data serves as the test set.

\subsection{Evaluate Event Clustering}
\label{subsec:eval-clustering}

We first evaluate the performance of our two-layer graph-based document clustering procedure for event extraction. We manually annotated a test dataset that consists of $3500$ news documents with ground-truth event labels, and compare our algorithm with the following methods:
\begin{itemize}
	\item \textbf{LDA + Affinity Propagation}: extract the 1000-dimensional LDA vector of each document, and cluster them by the Affinity Propagation clustering algorithm \cite{guan2011text}.
	\item \textbf{KeyGraph}: the original KeyGraph algorithm \cite{sayyadi2013graph} for document clustering, without the second-layer clustering based on document graphs and document-pair relationship classifier.
\end{itemize}

We use the homogeneity, completeness, and V-measure score \cite{rosenberg2007v} as the evaluation metrics of clustering results.
Homogeneity is larger if each cluster contains only members from a single class.  The completeness is maximized if all members of a ground true class are assigned to the same cluster.
The V-measure is the harmonic mean between homogeneity and completeness:
$
  \text{V-measure} = \frac{2 \times \text{homogenity} \times \text{completeness}}{\text{homogenity} + \text{completeness}}
$

\begin{table}
  \caption{Comparing different event clustering methods.}
  \label{tab:clusterResult}
  \begin{tabular}{llll}
    \toprule
    Algorithm & Homogeneity & Completeness & \text{V-measure}\\
    \midrule
    Our approach & $\mathbf{0.960}$ & $0.965$ & $\mathbf{0.962}$ \\
    KeyGraph & $0.554$ & $\mathbf{0.989}$ & $0.710$\\
    LDA + AP & $0.620$ & $0.947$ & $0.749$\\
    \bottomrule
  \end{tabular}
  \vspace{-5mm}
\end{table}

Table \ref{tab:clusterResult} shows that our approach achieves the best V-measure compared with other methods,  partly due to the fact that our method achieves the highest homogeneity score, which is $0.96$. This implies that most of the document clusters (events) we obtain are pure: each event only contains documents that talk about the same event. In comparison, the homogeneity for the other two methods is much lower. The reason is that we adopt two layers of graph-based clustering to group documents into events with more appropriate granularity. 

Yet, the completeness of our approach is a little bit smaller than that of KeyGraph, which is reasonable, as we further split the clusters with the second layer document-graph-based clustering supervised by the document-pair relationship classifier. Considering the significant improvement in homogeneity, the loss in completeness is ignorable. 

\subsection{Story Forest vs. Other Story Structures}
\label{subsec:eval-storytree}

We evaluate different event timeline and story generation algorithms on the large 3-month news dataset through pilot user evaluation. To make fair comparisons, the same preprocessing and event extraction procedures before developing story structures are adopted for all methods, with 261 stories detected from the dataset.
The only difference is how to construct the story structure given a set of event nodes.
We compare our online Story Forest system with the following existing algorithms:
\begin{itemize}
	\item \textbf{Flat Cluster (Flat):} this method clusters related events into a story without revealing the relationships between events, which approximates some previous works in TDT \cite{yang2002multi, allan1998line}.
	\item \textbf{Story Timeline (Timeline):} this method organizes events linearly according to the timestamps of events \cite{sayyadi2009event,sayyadi2013graph}.
	\item \textbf{Story Graph (Graph):} this method calculates a connection strength for every pair of events and connect the pair if the score exceeds a threshold \cite{yang2009discovering}.
	\item \textbf{Event Threading (Thread):} this algorithm appends each new event to its most similar earlier event \cite{nallapati2004event}. The similarity between two events is measured by the TF-IDF cosine similarity of the event centroids.
\end{itemize}

\begin{figure*}[t]
                        \centering
                        \subfigure[Percentage of incorrect edges]{
                \includegraphics[width=2.3in]{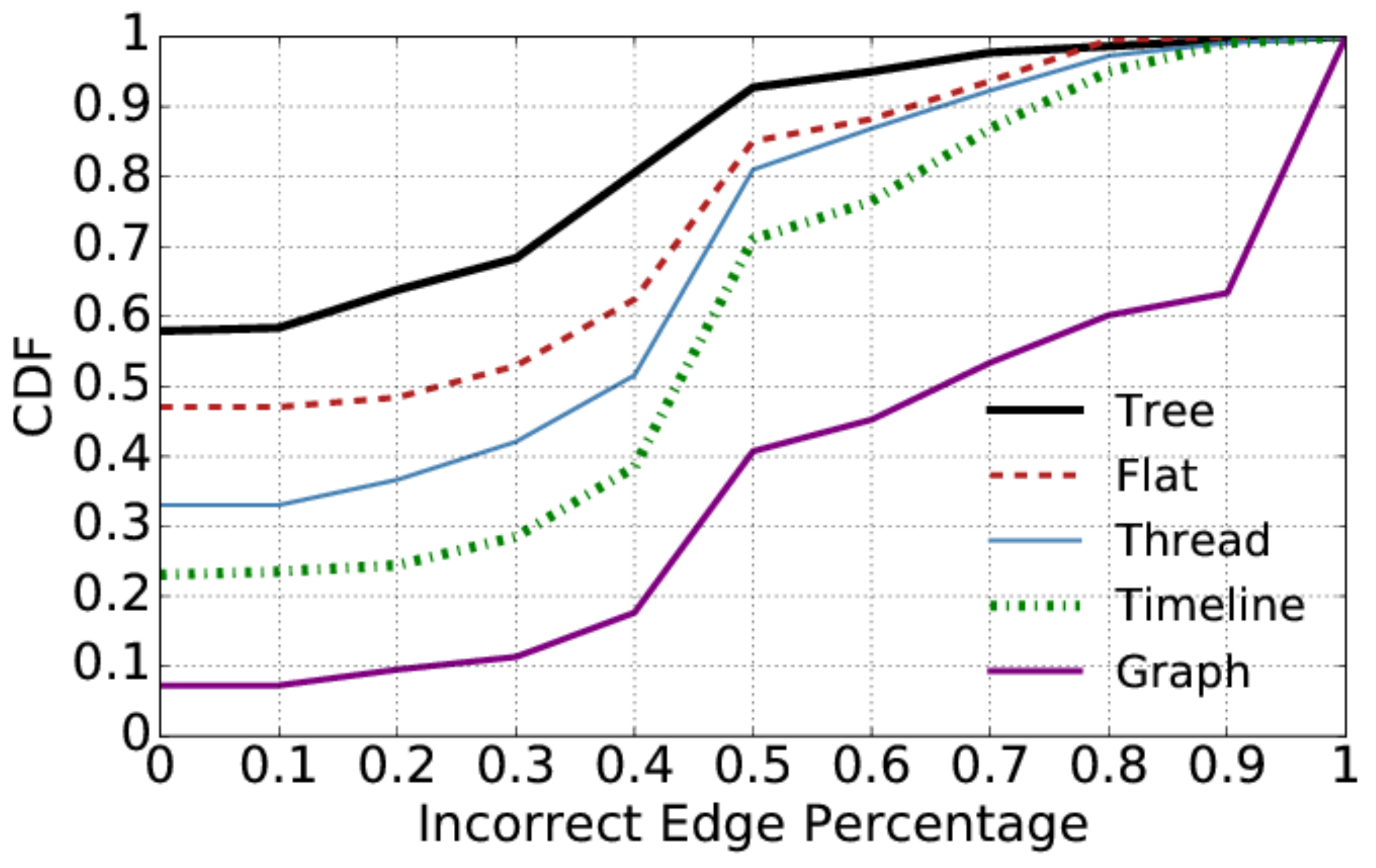}
                                \label{fig:edge}
                        }
                \hspace{-3mm}
                        \subfigure[Percentage of inconsistent paths]{
                \includegraphics[width=2.3in]{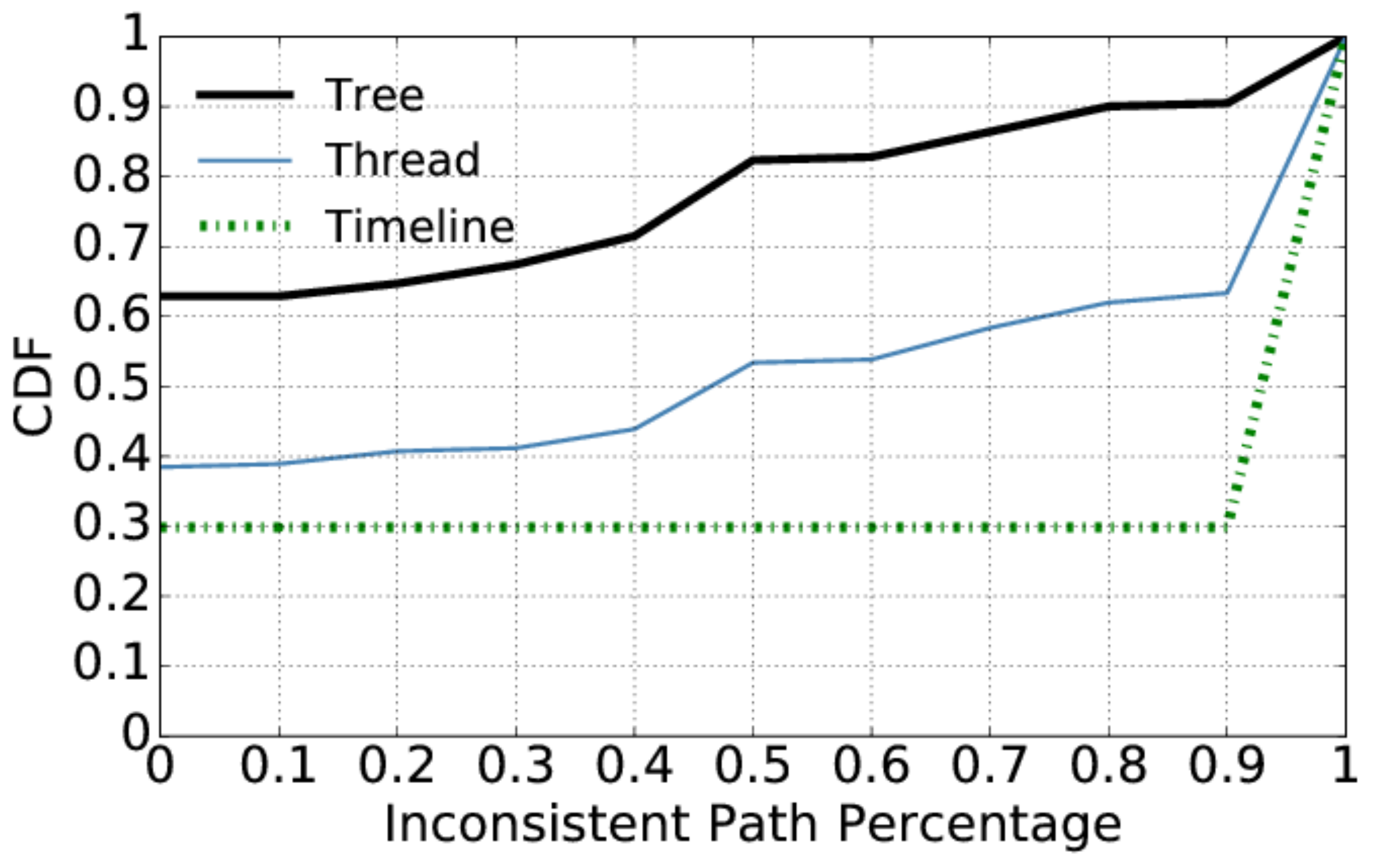}
                                \label{fig:path}
                        }
                        \hspace{-3mm}
                        \subfigure[Number of times rated as the most readable structure]{
                \includegraphics[width=2.3in]{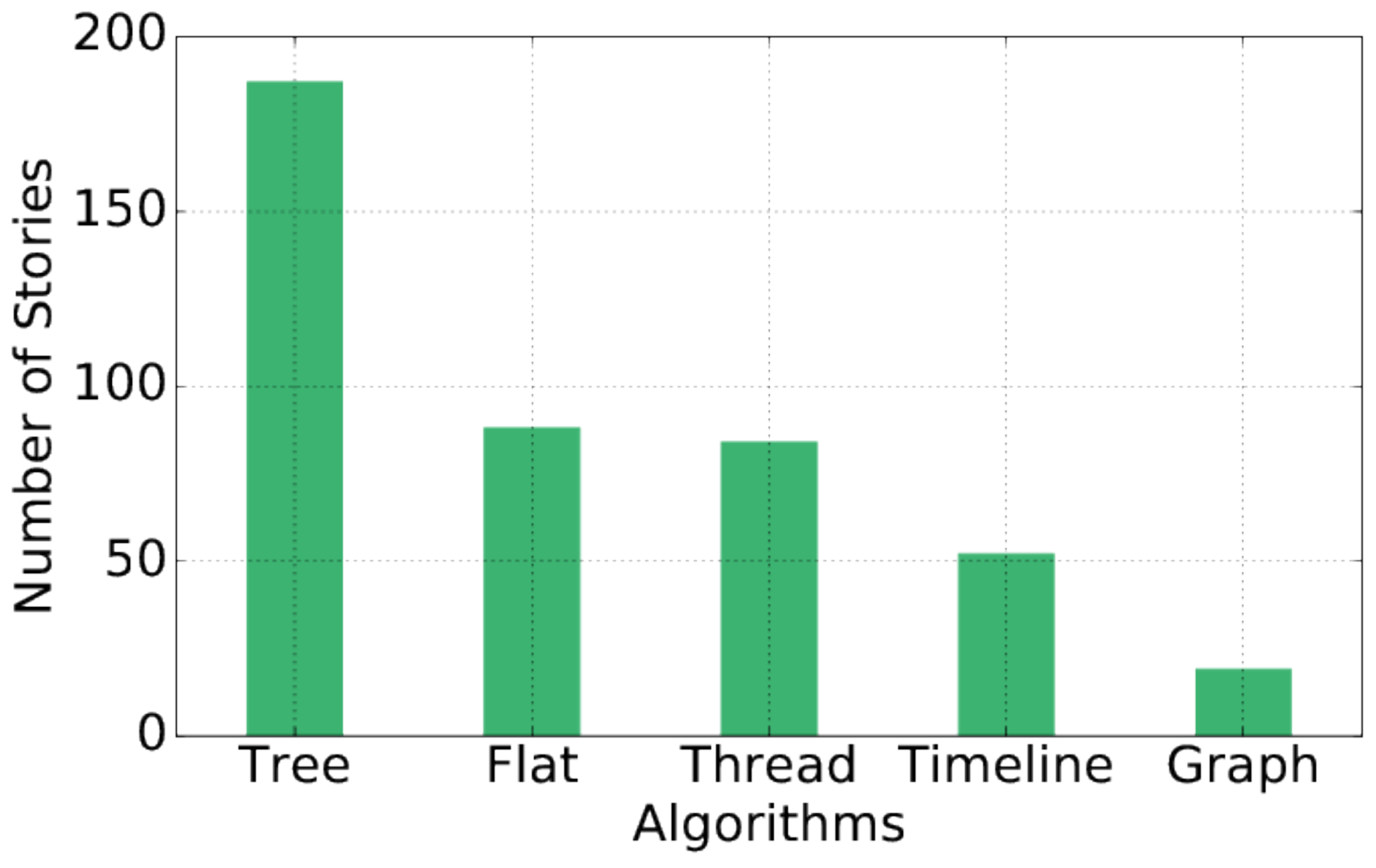}
                                \label{fig:whole}
                        }
                        \vspace{-3mm}
                \caption{Comparing the performance of different story structure generation algorithms.}
                \label{fig:compareAlgs}
\vspace{-4mm}
\end{figure*}

\begin{figure*}[t]
                        \centering
                        \subfigure[Histogram of the number of events in each story]{
                \includegraphics[width=2.3in]{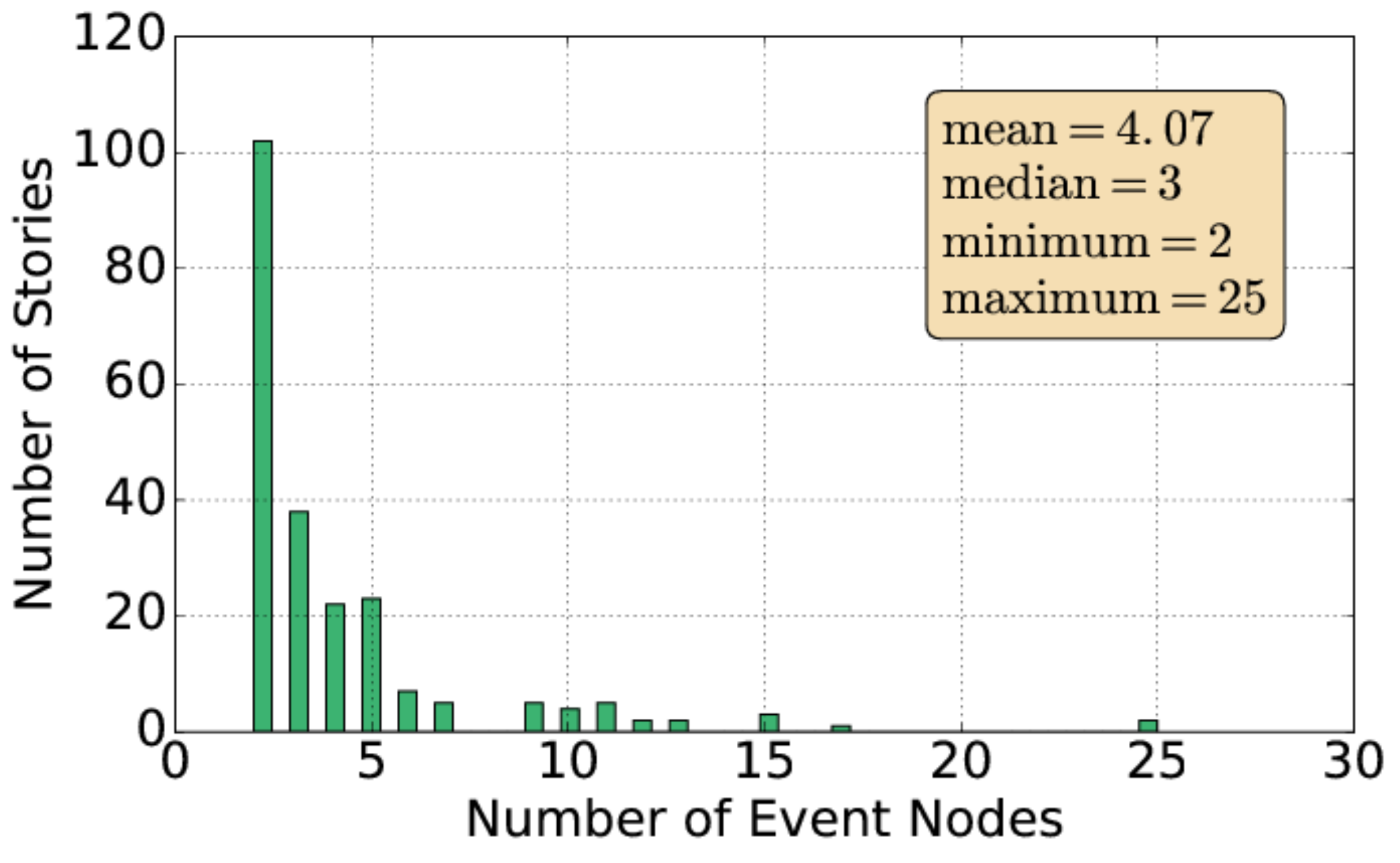}
                                \label{fig:num-event}
                        }
                \hspace{-3mm}
                        \subfigure[Histogram of the number of paths in each story]{
                \includegraphics[width=2.3in]{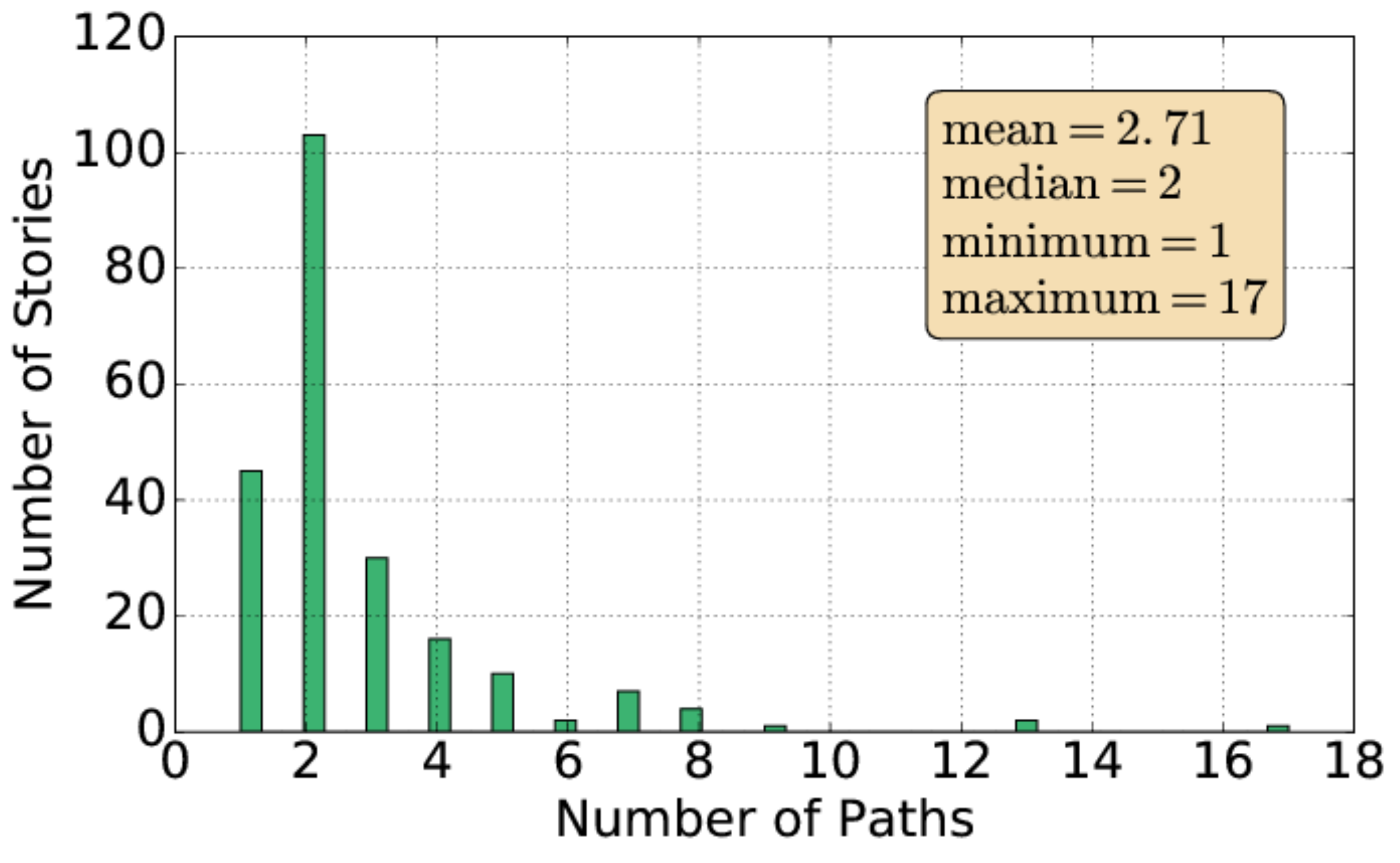}
                                \label{fig:num-path}
                        }
                        \hspace{-3mm}
                        \subfigure[Numbers of different story structures]{
                \includegraphics[width=2.3in]{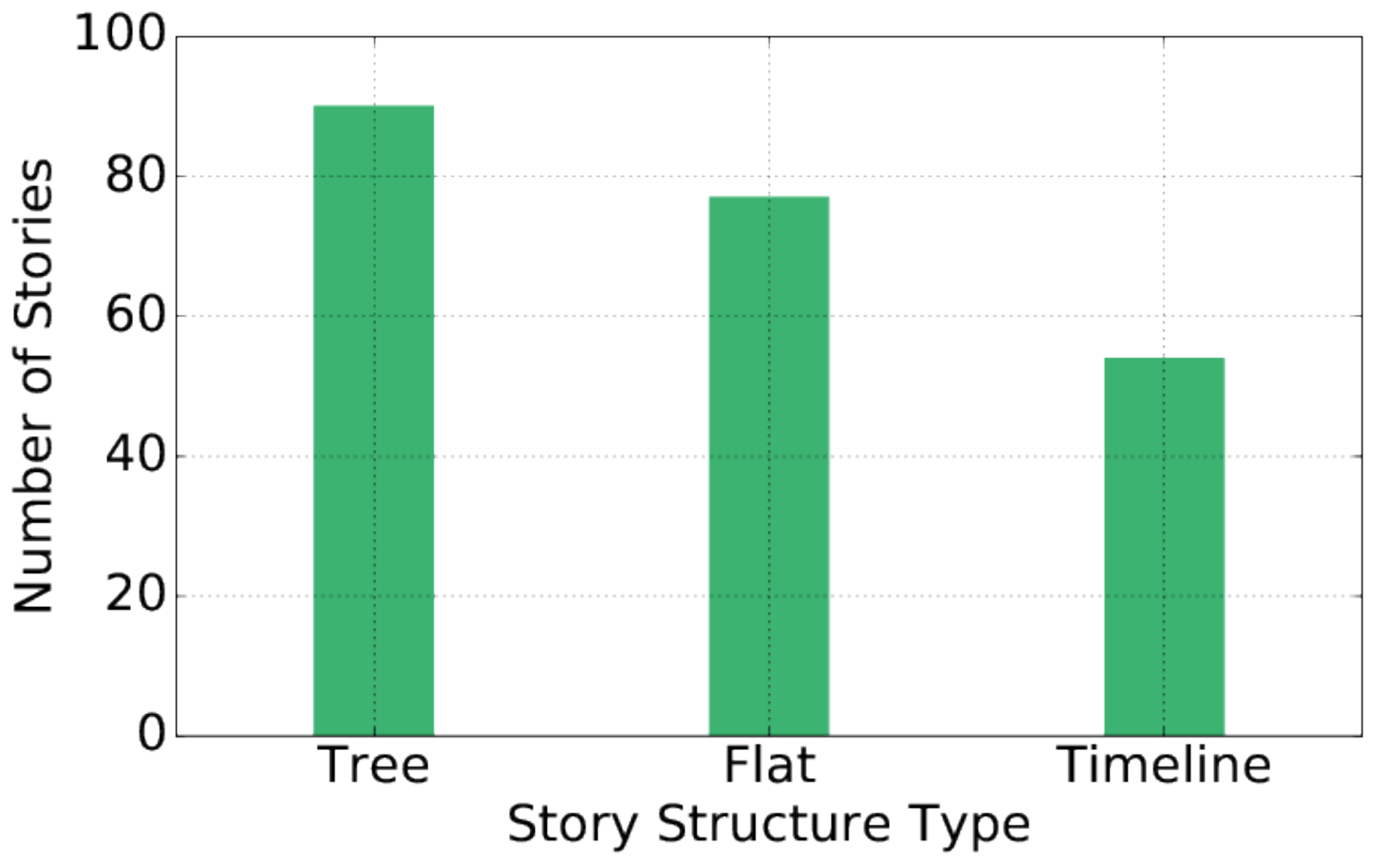}
                                \label{fig:story-type}
                        }
                        \vspace{-3mm}
                \caption{The characteristics of the story structures  generated by the  Story Forest system.}
                \label{fig:analysisTree}
\vspace{-4mm}
\end{figure*}

\begin{table}
  \caption{Comparing different story structure generation algorithms.}
  \label{tab:structureResult}
  \begin{tabular}{llllll}
    \toprule
     & Tree & Flat & Thread & Timeline & Graph\\
    \midrule
    Correct edges & $\mathbf{82.8\%}$ & $73.7\%$ & $66.8\%$ & $58.3\%$ & $32.9\%$ \\
    Consistent paths & $\mathbf{77.4\%}$ & $-$ & $50.1\%$ & $29.9\%$ & $-$\\
    Best structure & $\mathbf{187}$ & $88$ & $84$ & $52$ & $19$\\
    \bottomrule
  \end{tabular}
  \vspace{-3mm}
\end{table}

We enlisted $10$ human reviewers, including product managers, software engineers and senior undergraduate students, to blindly evaluate the results given by different approaches. Each individual story was reviewed by $3$ different reviewers. When the reviewers' opinions are different, they will discuss to give a final result. For each story, the reviewers answered the following questions for each of the $5$ different structures generated by different schemes:
\begin{enumerate}
	\item Do all the documents in each story cluster truly talk about the same story (\textit{yes} or \textit{no})? Continue if \textit{yes}.
	\item Do all the documents in each event node truly talk about the same event (\textit{yes} or \textit{no})? Continue if \textit{yes}.
	\item For each story structure given by different algorithms, how many edges correctly represent the event connections?
	\item For each story structure given by story forest, event threading and story timeline, how many paths from ROOT to any leaf node exist in the graph? And how many such paths are logically coherent?
	\item Which algorithm generates the structure that is the best in terms of revealing the story's underlying logical structure?
\end{enumerate}

Note that for question (3), the total number of edges for each tree equals to the number of events in that tree. Therefore, to make a fair comparison, for the story graph algorithm, we only retain the $n$ edges with the top scores, where $n$ is the number of events in that story graph.

We first report the clustering effectiveness of our system in the pilot user evaluation on the $3$-month dataset. Among the 261 stories, 234 of them are pure story clusters (\textit{yes} to question $1$), and furthermore there are 221 stories only contains pure event nodes (\textit{yes} to question $2$). Therefore, the final accuracy to extract events (\textit{yes} to both question $1$ and $2$) is $84.7\%$.

Next, we compare the output story structures given by different algorithms from three aspects: the correct edges between events, the logical coherence of paths, and the overall readability of different story structures. Fig.~\ref{fig:edge} compares the CDFs of incorrect edge percentage under different algorithms. As we can see, Story Forest significantly  outperforms the other $4$ baseline approaches.
As shown in Table~\ref{tab:structureResult}, for $58\%$ story trees, all the edges in each tree are reviewed as correct, and the average percentage of correct edges for all the story trees is $82.8\%$. In contrast, the average correct edge percentage given by the story graph algorithm is $32.9\%$. 

An interesting observation is that the average percentage of correct edges given by the simple flat structure is $73.7\%$, which is a special case of our tree structures.
This can be explained by the fact that most real-world breaking news that last for a constrained time period are not as complicated as a novel with rich logical structure, and a flat structure is often enough to depict their underlying logic.
However, for stories with richer structures and a relatively longer timeline, Story Forest gives better result than other algorithms by comprehensively considering the event similarity, path coherence and time gap, while other algorithms only consider parts of all the factors.

For path coherence, Fig.~\ref{fig:path} shows the CDFs of percentages of inconsistent paths under different algorithms.
Story Forest gives significantly more coherent paths: the average percentage of coherent paths is $77.4\%$ for our algorithm, and is $50.1\%$ and $29.9\%$, respectively, for event threading and story timeline. Note that path coherence is meaningless for flat or graph structure.

Fig.~\ref{fig:whole} plots overall readability of different story structures. Among the $221$ stories, the tree-based Story Forest system gives the best readability on $187$ stories, which is much better than all other approaches. Different algorithms can generate the same structure. For example, the Story Forest system can also generate a flat structure, a timeline, or a same structure as the event threading algorithm does. Therefore, the sum of the numbers of best results given by different approaches is bigger than $221$. It's worth noting that the flat and timeline algorithms also give $88$ and $52$ most readable results, which again indicates that the logic structures of a large portion of real-world news stories can be characterized by simple flat or timeline structures, which are special cases of story trees. And complex graphs are often an overkill. 

We further inspect the story structures generated by Story Forest. Fig.~\ref{fig:num-event} and Fig.~\ref{fig:num-path} plot the distributions of the number of events and the number of paths in each story tree, respectively. The average numbers of events and paths are $4.07$ and $2.71$, respectively. Although the tree structure includes the flat and timeline structures as special cases, among the $221$ stories, Story Forest generates $77$ flat structures and $54$ timelines, while the remaining $90$ structures generated are still story trees. This implies that Story Forest is versatile and can generate diverse structures for real-world news stories, depending on the logical complexity of each story.

\subsection{Algorithm Complexity and Overhead}
\label{subsec:complexity}

\begin{figure}[t]
		\hspace{7mm}
        \includegraphics[width=3.2in]{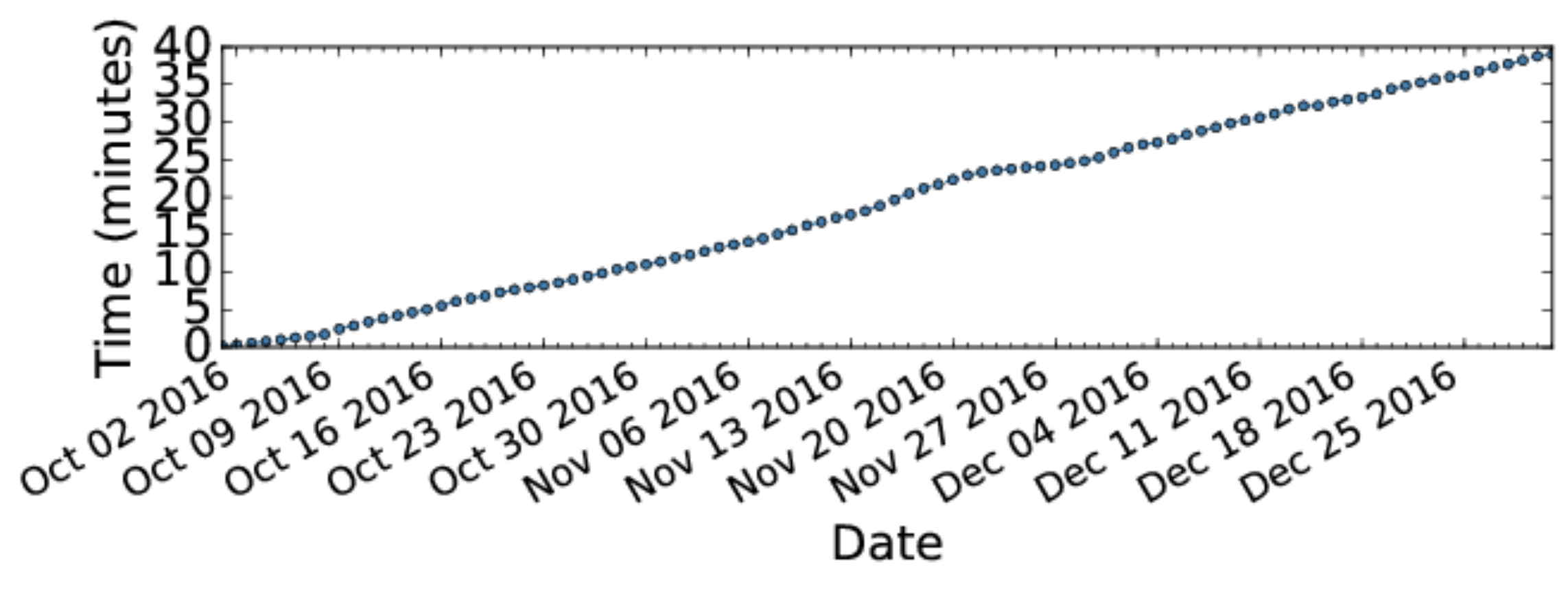}
        \vspace{-5mm}
        \caption{The running time of our system on the 3-month	 news dataset.}
        \label{fig:timeComplexity}
        \vspace{-5mm}
\end{figure}

In this section,  we discuss the complexity of each step in our system. For a time slot (e.g., in our case is one day), let $N_d$ be the number of documents, $N_w$ the number of unique words in corpora, note $N_w << N_d$, $N_e$ the number of different events, $N_s$ the number of different stories, and $N_k$ represents the maximum number of unique keywords in a document. 

As discussed in \cite{sayyadi2013graph}, building keyword graph requires $\mathcal{O}(N_d N_k + N_w^2)$ complexity, and community detection based on betweenness centrality requires $\mathcal{O}(N_w^3)$. The complexity of assigning documents to keyword communities is $\mathcal{O}(N_d N_k  N_e)$. So by far the total complexity is $\mathcal{O}(N_dN_k N_e + N_w^3)$. There exist other community detection algorithms requiring only $\mathcal{O}(N_w^2)$, such as the algorithm in \cite{radicchi2004defining}. Thus we can further improve efficiency by using faster community detection algorithms.

After clustering documents by keyword communities, for each cluster the average number of documents is $\sfrac{N_d}{N_e}$. The pair-wise document relation classification is implemented in $\mathcal{O}((\sfrac{N_d}{N_e})^2)$. The complexity of the next document graph splitting operation is $\mathcal{O}((\sfrac{N_w}{N_e})^3)$. Therefore, the total complexity is $\mathcal{O}(N_e((\sfrac{N_d}{N_e})^2 + (\sfrac{N_w}{N_e})^3))$. Our experiments show that usually $1 \leq \sfrac{N_d}{N_e} \leq 100$. Combining with $N_w << N_d$, the complexity is now approximately $\mathcal{O}(N_e)$.

To grow story trees with new events, the complexity of finding the related story tree for each event is of $\mathcal{O}(N_s  T)$, where $T$ is the history length to keep existing stories and delete older stories. If no existing related story, creating a new story requires $\mathcal{O}{(1)}$ operations. Otherwise, the complexity of updating a story tree is $\mathcal{O}(T \sfrac{N_e}{N_s})$. In summary, the complexity of growing story trees is $\mathcal{O}(N_eT (N_s + \sfrac{N_e}{N_s})) \approx \mathcal{O}(T N_e  N_s)$, as our experience on the Tencent news dataset shows that $1 \leq \sfrac{N_e}{N_s} \leq 200$. Our online algorithm to update story structure requires $\mathcal{O}(\sfrac{N_e}{N_s})$ complexity and  delivers a consistent story development structure, while most existing offline optimization based story structure algorithms require at least $\mathcal{O}((\sfrac{N_e}{N_s})^2)$ complexity and disrupt the previously generated story structures.

Fig.~\ref{fig:timeComplexity} shows the running time of our \textit{Story Forest} system on the $3$ months news dataset. The average time of processing each day's news is around $26$ seconds, and increases linearly with number of days. 
For the offline keyword extraction module, the processing efficiency is approximately $50$ documents per second. The performance of keyword extraction module is consistent with time and doesn't require frequently retraining. The LDA model is incrementally retrained every day to handle new words. For keyword extraction, the efficiency of event clustering and story structure generation can be further improved by a parallel implementation.

\section{Related Work}
\label{sec:related}

There are mainly two research lines that are highly related to our work: Text Clustering and Story Structure Generation.

The problem of text clustering has been well studied by researchers \cite{aggarwal2012survey,jing2005subspace,jing2010knowledge,guan2011text}. The most popular way is first extracting specific text features, such as TF-IDF, from documents, and then apply general clustering algorithms such as k-means. The selection of different feature and setting of algorithm parameters plays a key role in the final performance of clustering \cite{liu2005comparative}. There are also approaches which utilize the document keywords co-occurrence information to construct a keyword graph, and clustering documents by applying community detection techniques on the keyword graph \cite{sayyadi2013graph}. \cite{Mele2017Event} combines topic modeling, named-entity recognition, and temporal analysis to detect event clusters from news streams. \cite{Chakrabarti2010Evolutionary} proposed an evolutionary clustering framework to cluster data over time. A more comprehensive study of different text clustering algorithms can be found in \cite{aggarwal2012survey}.


The Topic Detection and Tracking (TDT) research spot news events and group by topics, and track previously spotted news events by attaching related new events into the same cluster \cite{allan1998line,allan2012topic,yang2009discovering,sayyadi2013graph}. However, the associations between related events are not defined or interpreted by TDT techniques. 
To help users capture the developing structure of events, different approaches have been proposed. \cite{nallapati2004event} proposed the concept of \textit{Event Threading}, and tried a series of strategies based on similarity measure to capture the dependencies among events. \cite{yang2009discovering} combines the similarity measure between events, temporal sequence and distance between events, and document distribution along the timeline to score the relationship between events, and models the event evolution structure by a directed acyclic graph (DAG). 

The above research works measure and model the relationship between events in a pairwise manner. However, the overall story consistency is not considered.
The \textit{Metro Map} model proposed in \cite{shahaf2013information} defines metrics such as coherence and diversity for story quality evaluation, and identifies lines of documents by solving an optimization problem to maximize the topic diversity of storylines while guarantee the coherence of each storyline. 
However, new documents are being generated all the time, and systems that are able to catch related news and update story structures in an online manner are desired.

As studies based on unsupervised clustering techniques \cite{yan2011evolutionary} perform poorly in distinguishing storylines with overlapped events \cite{hua2016automatical},
more recent works introduce different Bayesian models to generate storyline. However, they often ignore the intrinsic structure of a story \cite{huang2013optimized} or fail to properly model the hidden relations \cite{zhou2015unsupervised}. \cite{hua2016automatical} proposes a hierarchical Bayesian model for storyline generation, and utilize twitter hashtags to ``supervise'' the generation process. However, the Gibbs sampling inference of the model is time consuming, and such twitter data is not always available for every news stories.

\section{Conclusion}
\label{sec:conclude}

In this paper, we describe our experience of implementing \textit{Story Forest}, a news content organization system at Tencent, which is designed to discover events from vast streams of trending and breaking news and organize events in sensible story trees in an online manner. Our system is specifically tailored for fast processing massive amounts of breaking news data, whose story structures can most likely be captured by either a tree, a timeline or a flat structure.
We propose a two-layer graph-based document clustering algorithm to extract fine-grained events from vast long documents.
Our system further organizes the events into story trees with efficient online algorithms upon the arrival of daily news data. 
We conducted extensive performance evaluation based on 60 GB of real-world (Chinese) news data, although our ideas are not language-dependent and can easily be extended to other languages, through detailed pilot user experience studies.

Extensive results suggest that our clustering procedure is significantly more effective at accurate event extraction than existing algorithms. 
83\% of the event links generated by Story Forest are logically correct as compared to an accuracy of 33\% generated by more complex story graphs, demonstrating the ability of our system to organize trending news events into a logical structure that appeals to human readers.

\bibliographystyle{ACM-Reference-Format}
\bibliography{main} 

\end{document}